\documentclass[structabstract]{aa}
\usepackage{graphicx}
\usepackage{natbib}
\usepackage{color}
\usepackage{txfonts}
\def\degr{\hbox{$^\circ$}}

\begin{document}

\title{Correlation between technetium and lithium in a sample of oxygen-rich
  AGB variables\thanks{Based on observations at the Very Large Telescope of the
    European Southern Observatory, Cerro Paranal/Chile under Programme
    65.L-0317(A), and at the Calar Alto Observatory Centro Astron\'{o}mico
    Hispano Alem\'{a}n, Calar Alto, Spain}}

\author{Stefan Uttenthaler\inst{1}
  \and
  Thomas Lebzelter\inst{2}
}

\institute{
Instituut voor Sterrenkunde, K.\ U.\ Leuven, Celestijnenlaan 200D, 3000 Leuven,
Belgium\\
\email{stefan@ster.kuleuven.be}
\and
Department of Astronomy, University of Vienna,
T\"urkenschanzstra\ss e 17, 1180 Vienna, Austria\\
\email{lebzelter@astro.univie.ac.at}
}

\date{Received May 20, 2009; accepted October 27, 2009}


\abstract
{
The elements technetium and lithium are two important indicators of internal
nucleosynthesis and mixing in late type stars. Studying their occurrence and
abundance can give deep insight into the structure and evolution in the late
phases of the stellar life cycle.
}
{
The aims of this paper are: 1) to revisit the Tc content of a sample of
oxygen-rich asymptotic giant branch (AGB) variables and 2) to increase
the number of such stars for which the Li abundance has been measured to
provide constraints on the theoretical models of extra-mixing processes.
}
{
To this end, we analysed high-resolution spectra of 18 sample stars for the
presence of absorption lines of Tc and Li. The abundance of the latter was
determined by comparing the observed spectra to hydrostatic MARCS model
spectra. Bolometric magnitudes were established from near-IR photometry and
pulsation periods.
}
{
We reclassify the star V441~Cyg as Tc-rich, and the unusual Mira star R~Hya, as
well as W~Eri, as Tc-poor. The abundance of Li, or an upper limit to it, was
determined for all of the sample stars. In all stars with Tc we also detected
Li. Most of them have a Li content slightly below the solar photospheric value,
except for V441~Cyg, which has $\sim$1000 times the solar abundance. We also
found that, similar to Tc, a lower luminosity limit seems to exist for the
presence of Li.
}
{
We conclude that the higher Li abundance found in the cooler and higher
luminosity objects could stem from a production mechanism of Li operating on
the thermally pulsing AGB. The stellar mass might have a crucial influence on
this (extra mixing) production mechanism. It was speculated that the declining
pulsation period of R~Hya is caused by a recent thermal pulse (TP). While not
detecting Tc does not rule out a TP, it indicates that the TPs are not strong
enough to drive dredge-up in R~Hya. V441~Cyg, on the other hand, could either
be a low-mass, intrinsic S-star that produced its large amount of Li by
extra-mixing processes, or an intermediate-mass star
($M\gtrsim4\,\rm{M}_{\sun}$) undergoing Li production due to hot bottom
burning.
}

\keywords{stars: AGB and post-AGB -- stars: abundances -- stars: evolution
     -- stars: interiors -- stars: individual (R Hya, V441 Cyg)}
\authorrunning{Uttenthaler \& Lebzelter}
\titlerunning{Correlation between Tc and Li in AGB variables}
\maketitle

\section{Introduction}

The asymptotic giant branch (AGB) phase is the last stage of evolution
with nuclear burning for stars of low and intermediate mass
($1 - 8\rm{M}_{\sun}$). In the most luminous part of the AGB, the behaviour of
an AGB star is characterised by the so-called thermal pulses (TP), thermal
instabilities of the He shell accompanied by changes in luminosity,
temperature, pulsation period, and internal structure
\citep[see e.g.][for reviews]{Busso99,Herwig}. Between the repeated events of
explosive He-burning, heavy elements can be produced via the
``slow neutron capture process'' \citep[s-process, see e.g.][]{Wallerstein97}
in the region between the hydrogen- and the helium-burning shells. The
processed material may then be brought to the stellar surface by the convective
envelope that temporarily extends to these very deep layers. This mixing event
is called the third dredge-up (3DUP) and is the cause of the eventual
metamorphosis of an oxygen-rich M-star into a carbon-rich C-star.

The element technetium (Tc, $Z=43$) is a well-established diagnostic tool
in the study of the thermally pulsing AGB (TP-AGB) phase. Because of the short
half life time of the longest lived Tc isotope produced in the s-process
($^{99}$Tc, $\tau_{1/2}=2.1\times10^{5}$\,yr), any Tc that is detected in the
star's atmosphere must have been produced in-situ. This requirement has been
used to study the evolution of AGB stars ever since the pioneering discovery of
Tc absorption lines in the spectra of late type giants by \citet{Mer52b}.
Several classes of objects have been investigated for the presence of Tc in the
past \citep[e.g.][]{DW86,Lit87,VEJ99}. A key question in this research is if
and when in the evolution on the AGB, in terms of luminosity, the 3DUP sets in.
This question was addressed by \citet{LH99,LH03} for a sample of
solar-neighbourhood semi-regular pulsating AGB stars and for a larger sample of
Galactic disk AGB stars, including several Mira variables. Recently,
\citet{Utt07a} has investigated a sample of Galactic bulge AGB variables, for
which the distance and hence luminosity is more precisely known than for their
Galactic disk counterparts. The general conclusion of these works is that the
lower luminosity limit for 3DUP to set in, as predicted by theoretical
evolutionary models of the AGB, is consistent with the lowest luminosity
Tc-rich stars. However, being above this limit is not sufficient for a star to
have Tc in its atmosphere \citep{LH03,Utt07a}. In the present paper we revisit
the Tc content of a sample of oxygen-rich AGB stars, with important
consequences for a few individual stars.

The element lithium (Li, $Z=3$) is very fragile and temperature-sensitive.
Once it is mixed into stellar layers with temperatures exceeding
$3\times10^6$\,K, it is destroyed via proton captures to form helium. Its
fragility makes Li an important diagnostic tool for stellar evolution
\citep{Reb91}, and also has a high significance in the determination of
cosmological parameters \citep{Spi82,Kor06}. The standard model
\citep[e.g.][]{Michaud91} predicts a strong reduction in the Li surface
abundance after the first dredge up to values $\log \epsilon(\rm{Li}) < 1.5$,
or even less if depletion effects on the main sequence are considered. Most
giants for which a Li abundance has been determined
\citep[see a summary by][]{Michaud91} agree with this result.

Despite its fragility, Li can be produced rather than destroyed under certain
condition in stars via the so-called Cameron-Fowler or $^7$Be-transport
mechanism \citep[$^3$He($\alpha$,$\gamma$)$^7$Be($e^-$, $\nu$)$^7$Li; ][]{CF71}.
The production of Li is always connected to mixing mechanisms at or below the
bottom of the convective envelope, and requires $^{3}$He to be present in the
stellar atmosphere. One stage in stellar evolution where Li production occurs
is the luminosity bump of the red giant branch (RGB) when the H-burning shell
erases the molecular weight discontinuity left behind by the first dredge up. A
small fraction of giants in this phase are known to be very Li-rich
\citep{delaReza,CharBal}. The mixing process below the convective envelope has
been named extra-mixing or cool bottom processing \citep[CBP;][]{Was95}.
Its effects are also reflected in the carbon isotopic ratio measured from the
spectra of giant stars \citep{gil91}, as well as in other isotopic ratios
measured in meteoritic pre-solar grains of RGB or AGB origin \citep{Nol03}.

The physical cause of the extra-mixing below the convective envelope is
hitherto unknown. \citet{Egg06}, on the basis of a 3D simulation, find that
Rayleigh-Taylor instabilities below the convective envelope can develop from
the inversion of the mean molecular weight gradient induced by $^{3}$He burning.
Alternatively, \citet{ChaZah} suggest that the double-diffusive mechanism
called ``thermohaline instability'' should be at play. In principle, both these
phenomena might also occur on the AGB, though detailed models are just emerging
\citep{Can08}. Finally, \citet{Busso07} explore the possibility that circulation
of partially processed matter can be accounted for by the magnetic buoyancy
induced by a stellar dynamo operating on the RGB and on the AGB. That CBP is
probably active on the AGB has been recently demonstrated by \citet{Utt07b}
and \citet{Leb08}.

Among the AGB stars, those with carbon-rich chemistry (C-type, C/O$>$1 due to
the dredge-up of freshly synthesised $^{12}$C) have been studied more
intensively for their Li content than those with oxygen-rich chemistry (M-type,
C/O$<$1) in the past. This may partly have a historical reason because of the
early detection of carbon stars with a strong Li resonance line
\citep{McK40,Tor64}, but also an observational reason. Although the molecular
line blanketing mainly caused by the CN and C$_2$ molecules in the spectral
vicinity of the Li line in C-stars is strong, that of the TiO molecule in
oxygen-rich stars of the same effective temperature is even more severe. The
lower limit of Li abundance that can be measured is thus higher for O-rich
stars than for C-rich stars. In both cases, a Li abundance determination can
only be achieved with the help of spectral synthesis techniques including large
numbers of molecular lines. Thus, the counterparts to the catalogues of Li
abundance measurements in C-stars \citep{Den91,Bof93} include only relatively
small tables of M-type giants \citep{Luc82}. Even in this last work, a number
of the observed stars are supergiants and non-variable giants that are rather
in the RGB than in the AGB phase of evolution.

There are, however, M-type AGB stars that are known to be very rich in Li.
This type of star has been observed in the Magellanic Clouds \citep{Smith95},
and very recently also in the Milky Way galaxy \citep{GH07}. These objects are
intermediate-mass ($M \gtrsim 4 - 8\,\rm{M}_{\sun}$) TP-AGB stars with a
convective envelope penetrating the H-burning shell, thus nuclear-burning
occurs partly under convective conditions. This process is generally known as
hot bottom burning \citep[HBB;][]{Iben1973}. Under these conditions,
carbon, on the one hand, is converted into nitrogen, keeping the C/O ratio
below 1, and on the other hand $^7$Li is produced very efficiently by the
Cameron-Fowler mechanism. The Li abundance on the surface of HBB stars can be
higher than the initial abundance of the interstellar medium from which they
formed. We do not intend to deal with this kind of stars in the present paper,
but rather with the less massive AGB stars that have not (yet) dredged up
enough carbon to have C/O$>$1 and are not massive enough to remain O-rich due
to the operation of HBB. Nevertheless, by inspecting their bolometric
magnitudes, we will check whether our Li-rich objects belong to this group of
stars.

The correlation of Tc and Li was studied for a sample of Galactic S-type
(C/O$\simeq$1) stars by \citet{Van07}. \citet{Utt07b} present Li abundances
and their correlation with Tc in a homogeneous sample of O-rich AGB stars
located in the Galactic bulge. Here, we want to increase the sample of O-rich
AGB stars whose Li abundance has been measured, and study its correlation with
the presence of Tc in this type of stars. These measurements will be useful for
understanding the interaction between CBP mechanisms and 3DUP in AGB stars.

\smallskip
The paper is structured in the following way. In Sect.~\ref{samobs}, the sample
and the observations are described and bolometric magnitudes of the sample
stars derived; in Sect.~\ref{results} we present the analysis of the spectra,
results concerning the presence of Tc and the abundance of Li; in
Sect.~\ref{discussion} the results are discussed in view of a possible
enrichment mechanism of Li, and we take a closer look at two of the sample
stars; Sect.~\ref{summary} summarises the work.

\section{Sample and observations}\label{samobs}

The sample used in this study originally goes back to the search for Tc in
Galactic-disk AGB variables in the solar neighbourhood, based on parallaxes
measured with the Hipparcos satellite \citep{LH03}. Primary targets for that
study were those stars that had not been searched for Tc in the past and
objects without a definite decision on their Tc content. The original spectra
of that study are reanalysed in this work. An exception here is the star
\object{V441 Cyg}, whose observation and analysis was initiated by the work of
\cite{Van07}. A refined method for detecting Tc described in
Sect.~\ref{tcmethod} has led to some re-classifications in the sample, which we
describe in the present study.

Table~\ref{table_charac} lists the 18 objects studied in this project with a
number of important characteristics. The spectral type is a kind of ``mean''
value for what is found in the literature
\citep[using Vizier;][]{Vizier}\footnote{http://webviz.u-strasbg.fr/viz-bin/VizieR}.
The carbon star \object{TW Hor} is an exception in this sample. We include it
because it was observed in the same UVES run as most of the other sample stars.
According to the spectral type designations, the stars \object{W Cet} and
\object{V441 Cyg} have a notable enhancement of s-elements in their
atmospheres. V441~Cyg is described in the investigation of Galactic S-stars by
\citet{Van07}, who identify this star as the only one in their sample showing
Li in its spectrum, but not Tc. We come back to this star in
Sect.~\ref{sec_V441Cyg} where we challenge this classification.

\begin{table*}
\caption{Important characteristics of the sample stars. References for near
infrared data: 1: \citet{Catchpole}; 2: \citet{Fouquet}; 3: \citet{KH94};
4: \citet{K95}.}
\label{table_charac}
\begin{tabular}{lllrrcccc}
\hline\hline
Name             & Sp. type & Var. type & P [days] & $K$ [mag] & $J-K$ [mag] & Ref & $M_{\rm{bol}}$ & [12] -- [25] \\
\hline
\object{U Cet}   & M4       & M         & 234.76   & 2\fm73    & 1\fm11    & 1 & $-4\fm60$ & $-$1.14 \\
\object{W Cet}   & S7       & M         & 351.31   & 2\fm14    & 1\fm23    & 1 & $-5\fm07$ & $-$1.32 \\
\object{V441 Cyg}& M4S      & SRA       & 288.00   & 1\fm69    & 1\fm39    & 3 & $-5\fm66$ & $-$1.28 \\
\object{W Eri}   & M7       & M         & 376.63   & 1\fm78    & 1\fm42    & 1 & $-5\fm08$ & $-$0.90 \\
\object{S Gru}   & M7       & M         & 401.51   & 0\fm54    & 1\fm22    & 1 & $-5\fm35$ & $-$0.87 \\
\object{R Hor}   & M7       & M         & 407.60   & $-$0\fm88 & 1\fm24    & 1 & $-5\fm35$ & $-$0.92 \\
\object{T Hor}   & M5       & M         & 217.60   & 3\fm34    & 1\fm13    & 1 & $-4\fm44$ & $-$1.15 \\
\object{TW Hor}  & C5(N)    & SRB       & 158.00   & 0\fm19    & 1\fm46    & 3 & $-4\fm89$ & $-$1.03 \\
\object{R Hya}   & M7       & M         & 388.87   & $-$2\fm52 & 1\fm24    & 1 & $-5\fm28$ & $-$1.08 \\
\object{RU Hya}  & M6       & M         & 331.50   & 1\fm60    & 1\fm18    & 1 & $-5\fm07$ & $-$0.74 \\
\object{Y Lib}   & M5       & M         & 275.70   & 3\fm16    & 1\fm21    & 1 & $-4\fm72$ & $-$0.73 \\
\object{U Mic}   & M6       & M         & 334.29   & 1\fm80    & 1\fm22    & 1 & $-5\fm00$ & $-$0.69 \\
\object{Y Scl}   & M5       & SRB       &  79.36   & 0\fm35    & 1\fm27    & 2 & $-4\fm24$ & $-$0.94 \\
\object{T Tuc}   & M4       & M         & 250.30   & 2\fm96    & 1\fm18    & 2 & $-4\fm58$ & $-$1.08 \\
\object{ER Vir}  & M4       & SRB       &  55.00   & 1\fm44    & 1\fm10    & 4 & $-4\fm03$ & $-$1.41 \\
\object{EV Vir}  & M4       & SRB       & 120.00   & 1\fm55    & 1\fm15    & 4 & $-5\fm08$ & $-$1.34 \\
\object{RS Vir}  & M7       & M         & 353.35   & 1\fm15    & 1\fm31    & 1 & $-5\fm09$ & $-$0.56 \\
\object{S Vir}   & M7       & M         & 378.01   & 0\fm30    & 1\fm27    & 1 & $-5\fm21$ & $-$0.98 \\
\hline
\end{tabular}
\end{table*}

The variability type in column 3 of Table~\ref{table_charac} is taken from the
Combined General Catalogue of Variable Stars (CGCVS), version no.\ 4.2
\citep{Sam04}. Also the pulsation period in column 4 is taken from the CGCVS,
except for \object{Y Scl}, for which the period is taken from the All Sky
Automated Survey catalogue \citep{Poj05}, as well as \object{RS Vir} and
\object{S Vir}, for which the period is taken from \citet{Tem05}. A further
exception is \object{V441 Cyg}. Because of the importance of this star for the
further discussion of the results (see below), we made a check of the period
listed in the General Catalogue of Variable Stars (375\,d). We did a Fourier
analysis of the last 3000 days of the AAVSO archive data of this star
\citep{Henden09} using the program Period04 \citep{LB05}. We found a quite
well-expressed peak at 288\,d and a possible second period at 153\,d. A typical
visual light amplitude of 2\fm0 is found.

The near-IR photometry listed in columns 5 and 6 of Table~\ref{table_charac}
is taken from the references listed in column 7. The derivation of the
bolometric magnitude in column 8 is described in more detail in
Sect.~\ref{sect_mbol}. The IRAS colour listed in column 9 is defined as
$[12]-[25] = -2.5 \log(F_{12}/F_{25})$.

\smallskip
Most of the targets listed in Table~\ref{table_charac} were observed in July
2000 with the UVES spectrograph at ESO's Very Large Telescope under Programme
65.L-0317(A). The resolution of these spectra is
$R = \lambda/\Delta\lambda \simeq 50\,000$, and the wavelength coverage is
approximately 380 $-$ 490\,nm in the ``blue arm'', and 667 $-$ 1000\,nm in
the ``red arm'' of UVES. These observations are described in \citet{Utt07a},
and the spectra are analysed for Tc in \citet{LH03}. We thus refer to these
papers for more details. The only exceptions to this are the stars
\object{W Cet}, which is not included in \citet{LH03}, and \object{V441 Cyg},
which was observed for the present study in service mode with the FOCES
spectrograph \citep{Pfe98} mounted on the 2.2\,m telescope of the Calar Alto
Observatory in southern Spain on 23 June 2008. The observations were carried
out using the CCD chip with 15\,$\mu$m pixel size and a slit width of
100\,$\mu$m, which subtends 1\farcs15 on the sky. The chosen setup yields
a spectral resolving power of $R \simeq 65\,000$. The exposure time was two
times 1800\,s. Because of the high dynamic range of FOCES and the resulting
reduction problems with the ``red'', ``green'', and ``blue'' flats, the spectra
were reduced with a custom-made IDL routine. The signal-to-noise ratio (SNR) of
the spectra is estimated to be in the range 6 $-$ 10 at the wavelength of the
Tc lines, and $\gtrsim$100 at the wavelength of the 670.8\,nm Li
line\footnote{All wavelengths in this paper are expressed in nm in air.}. As
shown by \citet{Utt07a}, an SNR of 6 -- 10 is sufficient to decide on the
presence of Tc based on a flux-ratio method (see next section), as long as the
spectral resolution is high enough.

The general appearance of the spectra is dominated by strong bands of the
TiO molecule (except of the carbon star \object{TW Hor}, of course), clearly
verifying their oxygen-rich nature. Some of the stars also show significant
bands of the ZrO molecule, indicating that some s-process enrichment has taken
place. Only stars that are identified as Tc-rich (see below) also show bands of
ZrO, and conversely no Tc-poor star shows any clear sign of ZrO absorption
bands. Thus, none of our sample stars is extrinsically enriched in s-process
elements by, e.g., mass transfer from a binary companion. However, the
absorption bands of the LaO molecule cannot be clearly identified in any of our
sample stars, as is sometimes the case in S-type stars.

\subsection{Bolometric magnitude determination}\label{sect_mbol}

To derive absolute $K$-magnitudes, we used the log$P-K$ relations from
\citet{Ita04}. For stars classified as Miras we used Ita's relation C
including their separation for targets with $J-K < 1.4$ and $J-K > 1.4$. For
semi-regular variables we used sequence B or C'. Naturally, this approach
includes some uncertainties. In their Table 3, \citet{Ita04} list a typical
width of about 0\fm2 of these sequences. This, however, is much less than the
uncertainties of absolute $K$-magnitudes derived from parallax measurements,
which are only available for a small number of stars in our sample. A larger
error comes from the uncertainty of attributing a star to a given pulsation
sequence. Several authors agree \citep{Wood99,KB03,Lebzelter05} that large
amplitude variables, classically dubed Miras, are found on sequence C, which
corresponds to fundamental mode pulsation. The risk of a misclassification will
thus be low for the Miras. It is much more difficult to attribute the
inhomogeneous group of the five semi-regular sample stars correctly. They can be
located on any sequence from A to C, and the definition of sequences A, B, and
C' is less clear. As selection criterion, we chose the light amplitude following
the extensive analysis of \citet{KB03} for the LMC. As on average the amplitude
becomes smaller from sequence C' to A, we attributed the non-Miras
\object{V441 Cyg}, \object{Y Scl}, and \object{EV Vir} to sequence C',
\object{TW Hor} and \object{ER Vir} to sequence B. It is clear that this
classification and the derived luminosities have to be taken with some caution.
However, because of the small separation of sequences B and C' compared to the
difference to sequence C, any misclassification will not change the overall
result substantially.

To derive absolute bolometric magnitudes from the absolute $K$-magnitudes,
we need to apply a bolometric correction. We adopt the approach outlined
in \citet{Mekul} using two different relations between blackbody fits and
$J-K$ depending on the $K-L$ colour. The inclusion of a second near infrared
colour allows for much better treatment of stars at the cool end of the AGB
than other formulas found in the literature. Details will be given elsewhere
(Kerschbaum et al., in preparation). 

We used near-infrared colours from the literature (see
Table~\ref{table_charac}). The $J$ and $K$ band magnitudes were taken from
\citet{Catchpole}, \citet{Fouquet}, \citet{KH94} and \citet{K95}, respectively.
All infrared measurements were converted to the 2MASS system using the
relations given in \citet{Carpenter}. Where more than one measurement was found
we used an average value. While the $K$-magnitudes used to derive the absolute
bolometric magnitudes were calculated from a star's period, the bolometric
correction is based on individual near infrared measurements and thus affected
by stellar variability. The extensive material presented by \citet{Catchpole}
allow estimation of the typical uncertainty of the bolometric correction from
the variations in the $J-K$ colour. For Miras, this uncertainty can be up to
0\fm5. In a few cases no $K-L$ colour was available. We then used an average of
the two relations. In all cases the error introduced by this is less than
0\fm1. Interstellar extinction has not been taken into account for these rather
nearby objects.

\begin{figure}
  \centering
  \includegraphics[width=\linewidth,bb=6 3 534 381]{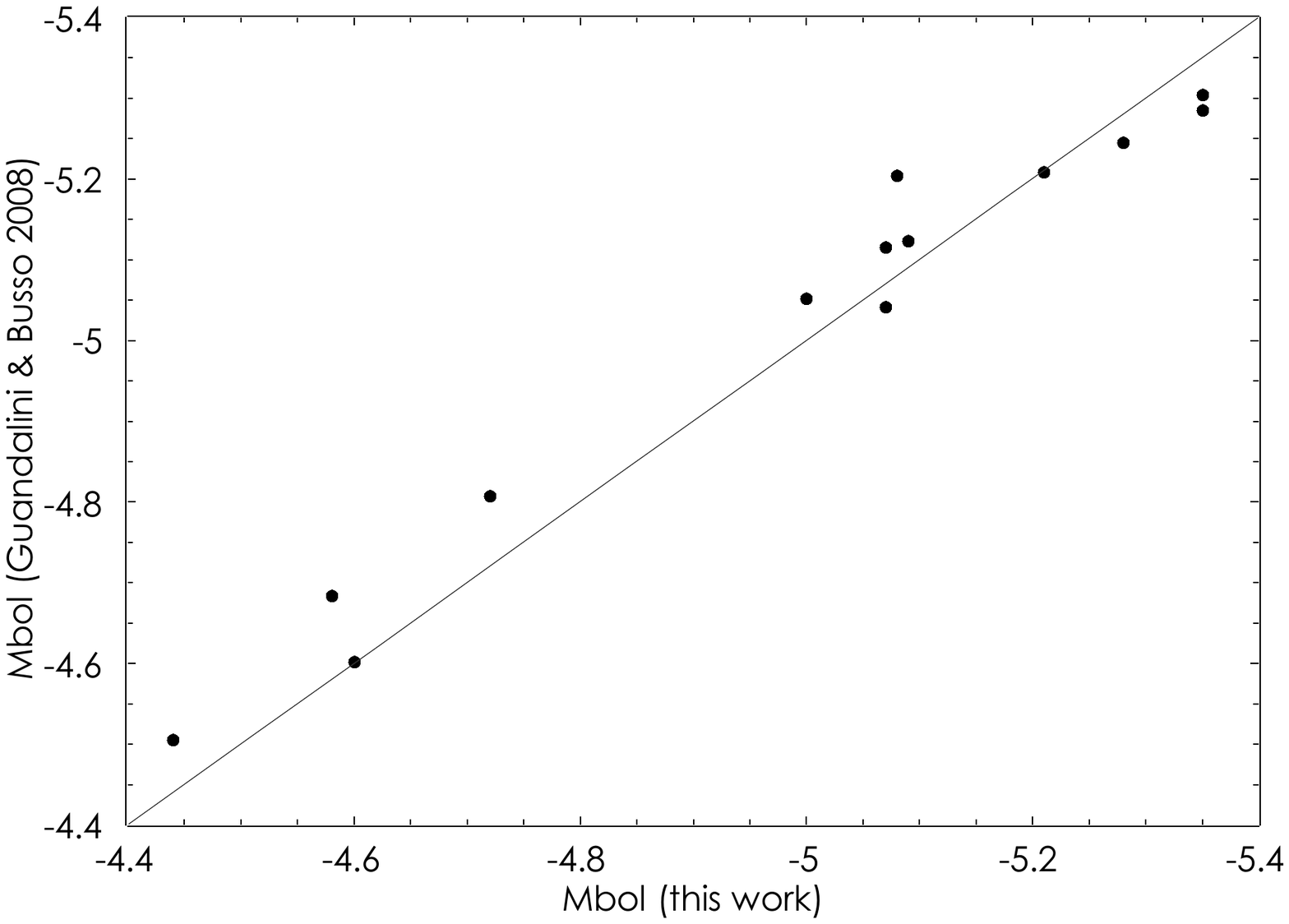}
  \caption{Comparison of $M_{\rm{bol}}$ values derived by out method compared
  with values using the formula given in \citet{Gua08}. The full line indicates
  the 1:1 relation. Good agreement can be seen.}
  \label{mbolcomp}
\end{figure}

Recently, \citet{Gua08} have published a new set of bolometric magnitudes for
mass-losing AGB stars. As a result they give a $\log P - M_{\rm{bol}}$ relation
for Miras (see their Fig.\ A.1). Using this relation we calculated
$M_{\rm{bol}}$ for the Miras in our sample. In Fig.~\ref{mbolcomp} we compare
the results derived from \citet{Gua08} with our values of the absolute
bolometric magnitude for the stars in common, showing good agreement between
the two approaches.

\section{Analysis and results}\label{results}

\subsection{Analysis with respect to Tc}\label{tcmethod}

The radial velocity (rv) is the first quantity that we determined from our
spectra. It was measured from the TiO band head at 705.4\,nm using a
cross-correlation technique, with a synthetic MARCS spectrum as template.
Sect.~\ref{sect_param} gives more details on the used MARCS models. We chose
this TiO feature because the SNR of the observed spectra is much higher at this
wavelength than at the wavelength of the Tc lines, and the TiO features are
formed in the stellar atmosphere close to the atomic (Tc) lines. For
\object{TW Hor}, the rv was measured from atomic lines in the vicinity of the
Tc lines, due to the absence of TiO bands in carbon stars. The precision of the
rv measurement is of the order of 1\,km\,s$^{-1}$ or better. The differences to
literature values are usually small (a few km\,s$^{-1}$), and can be fully
explained by rv variations caused by the stellar pulsation.

\begin{figure}
  \centering
  \includegraphics[width=\linewidth,bb=82 370 536 700]{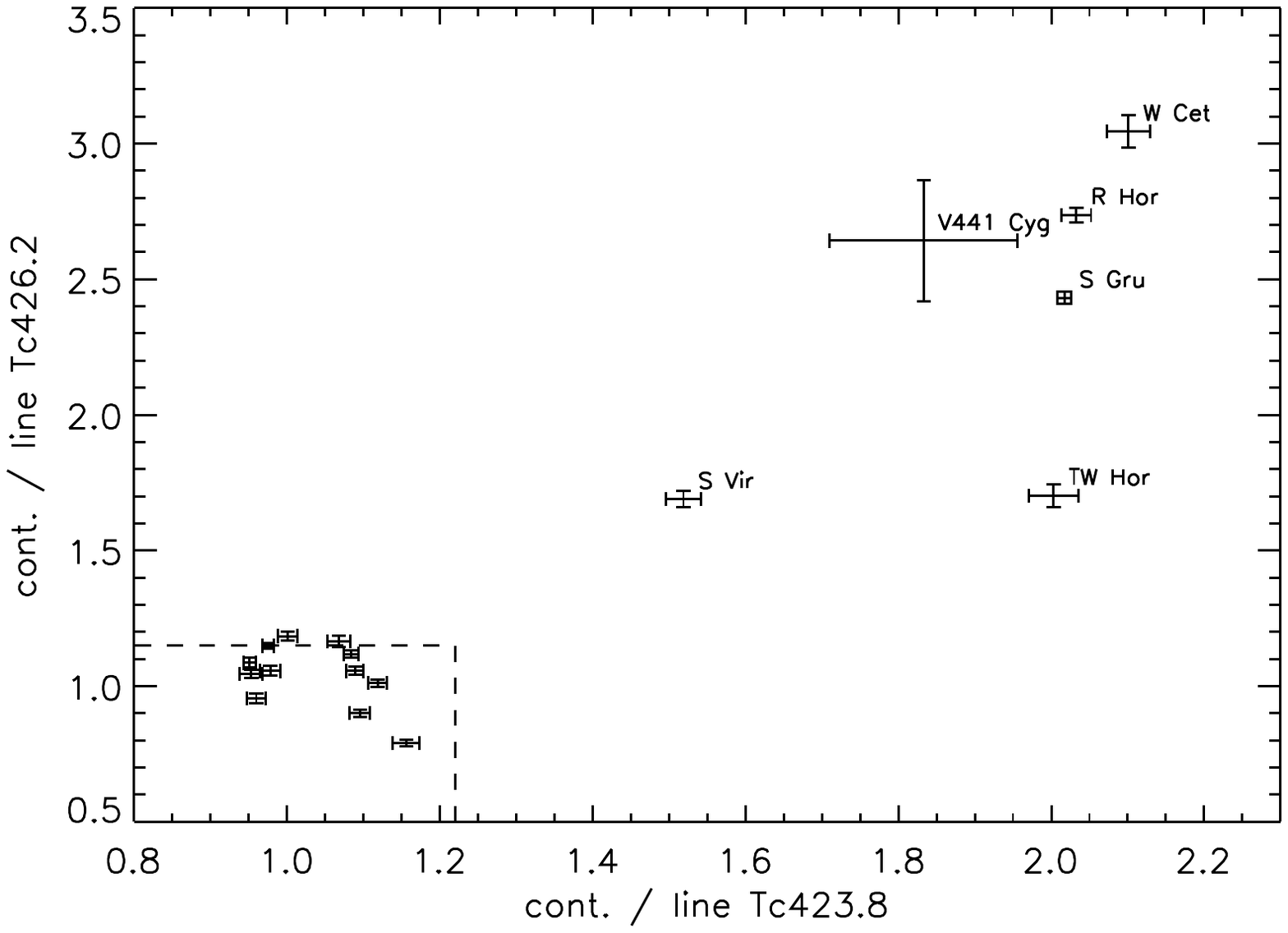}
  \caption{Ratio of the continuum flux to line core flux for the two Tc lines
  at 423.819 and 426.227\,nm. The quasi-continuum points were taken in the range
  423.940 $-$ 423.970\,nm and 426.130 $-$ 426.150\,nm, respectively, the line
  core flux was measured in the range 423.805 $-$ 423.835\,nm and 426.220 $-$
  426.240\,nm, respectively. While the Tc-poor stars gather around ratios 1.0
  (lower left corner), the Tc-rich stars are displaced to the upper right
  by the Tc line absorption. The dashed lines represent the upper limits of the
  ratios found for the Tc-poor bulge stars analysed in \citet{Utt07a}.}
  \label{line_ratio}
\end{figure}

We corrected the observed spectra by the measured rv shift to check them for
the Tc lines using the flux-ratio method introduced by \citet{Utt07a}. In this
method, the mean flux in a small wavelength interval around a quasi-continuum
point close to the Tc line is divided by the mean flux in a small wavelength
interval centred on the Tc line itself. This ratio can be calculated for all of
the seven identified Tc lines (398.497, 403.163, 404.911 and 409.567, 423.819,
426.227, and 429.706\,nm), and plots of the ratios from different lines clearly
separate the Tc-poor from the Tc-rich stars. If the Tc lines are absent, the
ratio will be close to 1.0, but significantly more than 1.0 if they are
present. Figure~\ref{line_ratio} shows such a plot for the lines at 423.819 and
426.227\,nm for our sample stars; plots involving other Tc lines are very
similar and reveal the same Tc-rich stars. The Tc-rich stars that clearly
separate from the Tc-poor ones are identified in the plot with their name. The
error bars on the symbols in Fig.~\ref{line_ratio} were directly derived from
the SNR of the observed spectra.

We want to briefly explain the advantages of the flux-ratio method over
previous methods of deciding on the presence of Tc. The method previously
applied uses the central wavelength of the Tc line blend at 429.706\,nm
\citep[e.g.][and references therein]{Lit87,VEJ99}. This Tc line is blended with
a line of chromium. If Tc is present, the central wavelength is shifted to a
somewhat longer wavelength than in the absence of Tc. However, the shift is not
large, thus the separation between Tc-poor and Tc-rich stars is not very large
either. Every uncertainty of the local wavelength scale has a direct impact on
the central wavelength of the blend, thus adding uncertainty to the
classification with respect to Tc. Due to the stellar pulsation and the
velocity fields in the atmosphere, lines formed at different layers have a
slightly different rv shift. This shift also depends on the pulsation phase at
which the observations were done. The result is that intermediate cases are
classified as ``Tc doubtful'', ``Tc possible'', and ``Tc probable''.
Figure~\ref{line_ratio} shows that, with our flux-ratio method, the gap between
Tc-poor and Tc-rich stars is so large compared to the uncertainty that no room
is left for these intermediate cases. The uncertainty of the wavelength scale
has only a very small impact on this method. We can thus very clearly decide on
``Tc yes'' or ``Tc no''. We conclude that the flux-ratio method delivers more
reliable results than the central wavelength method.

Table~\ref{table_measure} summarises the radial velocity shifts measured from
the spectra and the updated Tc classification. With respect to previous
classifications, there are three changes: \object{V441 Cyg} is Tc-rich
\citep[][Tc-poor]{Van07}, \object{W Eri} is Tc-poor
\citep[][Tc possible]{LH03}, and \object{R Hya} is Tc-poor as well
\citep[][Tc-rich]{LH03}. 

\begin{table}
\caption{Quantities measured from the spectra: radial velocity, Tc content,
effective temperature, and Li abundance of the sample stars.}
\label{table_measure}
\begin{tabular}{lrccr}
\hline\hline
Name             & rv [km\,s$^{-1}$] & Tc? & $T_{\rm{eff}}$ & $\log\epsilon(\rm{Li})$ \\
\hline
\object{U Cet}   & $-19.0$  & no  & 3300 & $\lesssim0.0$ \\
\object{W Cet}   & $ +5.8$  & yes & 3100 & +0.8          \\
\object{V441 Cyg}& $-61.7$  & yes & 3400 & +4.6          \\
\object{W Eri}   & $+17.8$  & no  & 2950 & +0.4          \\
\object{S Gru}   & $-18.4$  & yes & 2900 & +0.8          \\
\object{R Hor}   & $+56.0$  & yes & 2900 & +0.6          \\
\object{T Hor}   & $+44.9$  & no  & 3300 & $\lesssim0.0$ \\
\object{TW Hor}  & $+14.5$  & yes & 3000 & $-$1.5        \\
\object{R Hya}   & $-15.7$  & no  & 3100 & $\lesssim0.0$ \\
\object{RU Hya}  & $ +0.4$  & no  & 3300 & $\lesssim0.0$ \\
\object{Y Lib}   & $ +9.3$  & no  & 3100 & +0.9          \\
\object{U Mic}   & $-61.0$  & no  & 2900 & +0.6          \\
\object{Y Scl}   & $+30.0$  & no  & 3200 & +1.0          \\
\object{T Tuc}   & $-37.9$  & no  & 3300 & $\lesssim0.0$ \\
\object{ER Vir}  & $ +8.5$  & no  & 3500 & $\lesssim0.0$ \\
\object{EV Vir}  & $-44.5$  & no  & 3500 & $\lesssim0.0$ \\
\object{RS Vir}  & $-15.9$  & no  & 2900 & +1.0          \\
\object{S Vir}   & $ +7.2$  & yes & 2800 & +0.5          \\
\hline
\end{tabular}
\end{table}

\subsection{Analysis with respect to Li}

The abundance of Li in stars is usually measured from the strong resonance
doublet of neutral Li at 670.8\,nm. The line strength is quite sensitive to
the effective temperature of the star, thus a good knowledge of this stellar
parameter is required for reliable abundance determination. In the following we
describe in some detail how the temperature determination was accomplished for
the present sample of stars.

\subsubsection{Effective temperature determination}
\label{sect_param}

For determining the effective temperatures of the sample stars, we applied a
$\chi^2$ minimisation method to band heads of the TiO molecule using a small
grid of synthetic model spectra. In the optical range, TiO bands are the
dominant spectral features of cool oxygen-rich giants, and they are very
sensitive to the stellar temperature. We thus followed a procedure similar to
\citet{GH07}, who apply this method to measure Li abundances in massive
Galactic AGB (OH/IR) stars. We used the same modified $\chi^2$ test as these
authors to fit model data $Y_{synth,i}$ to observed data $Y_{obs,i}$, which is
expressed in the form
\begin{equation}
\chi^2 = \sum_{i=1}^{N} \frac{\left[ Y_{obs,i} - Y_{synth,i}(x_{1}\dots x_{M})\right]^2}{Y_{obs,i}},
\end{equation}
with N the number of data points and M the number of free parameters. The
synthetic spectra are based on COMARCS atmospheric models. COMARCS is a revised
version of MARCS \citep{Joe92} with spherical radiative transfer routines from
\citet{Nordlund} and new opacity data from the COMA program \citep{Ari00}.
The only model parameters that we varied in this procedure were the effective
temperature and the metallicity. The other stellar parameters have a very
limited impact on the spectral appearance, hence on the results of this
analysis \citep[see also][for a more detailed discussion]{GH07}.
For the surface gravity $\log g$\,[cm\,s$^{-1}$], micro-turbulent velocity
$\xi$, and the mass $M$, we chose values typical for red giant stars of 0.0,
3\,km\,s$^{-1}$, and 1\,M$_{\sun}$. Since the precise C/O ratio cannot be
determined from our optical data (near-IR spectra would be required to do so),
we fixed it at the solar value of 0.48, although in principle it might deviate
from that value, particularly for the Tc-rich sample stars. Once the C/O ratio
approaches 1, the chemistry and thus the dominant features in the spectrum will
change significantly. Two stars of our sample are classified as MS or S-type
(\object{W Cet} and \object{V441Cyg}), and only for these objects do we expect
that the assumption of a constant C/O might deteriorate the results to some
degree. We come back to this point in the next section.

We chose three metallicity values for the grid of model atmospheres: 0.3, 1.0,
and 1.6 times the solar metallicity. The temperature ranges between 2600 and
3600\,K in steps of 100\,K, with intermediate steps at 2950 and 3150\,K for the
solar metallicity models. From this grid of model atmospheres, synthetic
spectra were calculated in the two spectral pieces 668 -- 674\,nm (covering the
Li line at 670.8\,nm and the TiO $\gamma$(1,0)Rc and TiO $\gamma$(1,0)Rb
band-heads) and 700 -- 710\,nm (covering the strong TiO $\gamma$(0,0)Ra
band-head at $\sim 705$\,nm and the TiO $\gamma$(0,0)Rb band head). Atomic
lines from the VALD database \citep{Kup99} and TiO lines from the list of
\citet{Sch98} were taken into account in the spectral synthesis. Other
molecular line data of less impact on the spectral ranges under consideration
were also taken into account; these data are summarised in Table~1 of
\citet{Cristallo}.

Prior to the $\chi^2$ minimisation fitting, the model spectra were convolved
with a Gaussian to a resolution of $R = \lambda/\Delta\lambda= 50\,000$ (except
for V441~Cyg, for which the spectra are convolved to $R = 65\,000$). An
additional macro-turbulence of 3\,km\,s$^{-1}$ was assumed. This proves an
acceptable  value for all except two stars, namely V441~Cyg and Y~Scl, for
which a higher value of 6\,km\,s$^{-1}$ had to be adopted to make a
satisfactory fit to the spectra.

Although the metallicities adopted for the atmospheric model grid cover a quite
wide range, it turned out that the temperature at which $\chi^2$ becomes
minimal does not strongly depend on the metallicity. The $\chi^2$ minimum for
the spectral piece including the Li line is quite flat for many stars, thus
the adopted temperature was mostly fixed by the second piece between
700 -- 710\,nm. For the spectral synthesis to determine Li abundances (see next
section), we therefore adopted the model with solar metallicity, with a
temperature that satisfactorily fits both spectral pieces. There is some
indication in the literature that solar metallicity is a good assumption for
solar neighbourhood AGB stars \citep[e.g.][]{Van02,GH07}. After finishing this
procedure, we found that the adopted effective temperature correlates well with
the $(J - K)$ colour, as can be expected: cooler stars have higher (redder)
$(J - K)$ colour than hotter stars. The adopted temperatures are summarised in
the fourth column of Table~\ref{table_measure}. We estimate that the precision
of the temperature determination is $\pm 100$\,K.

\subsubsection{Li abundances}

With this temperature, spectral synthesis calculations of the Li line doublet
were performed to measure the Li abundance, using the MARCS models with solar
metallicity. The $\log gf$ values from the VALD database were adopted for
the 670.8\,nm Li doublet ($\log gf = -0.009$ and $-0.309$ for the component at
the shorter and longer wavelength, respectively). The Li abundance was varied
in steps of 0.1\,dex, and the observed-minus-calculated flux at the wavelength
of the Li line was monitored to determine the Li abundance in the atmosphere of
each sample star. The abundance that left practically a zero flux difference
was finally adopted. The last column of Table~\ref{table_measure} gives the
measured Li abundances on the $\log\epsilon$ scale
($\log \epsilon(\rm{Li}) = \log N(\rm{Li})/N(\rm{H}) + 12$). The solar
photospheric Li abundance on this scale is $\log \epsilon(\mathrm{Li}) = +1.1$.
We also used our model spectra to verify that the observed absorption around
the Li line cannot come from a line of cerium, which is observed to be a
``substitute'' for Li in the hotter barium stars and post-AGB stars
\citep{Lam93,Rey02}.

From the uncertainty of the temperature, the stellar parameter with the
strongest influence on the Li line strength, we estimate an abundance
uncertainty of $\pm 0.4$\,dex. Taking the uncertainties of the other
stellar parameters into account, which are harder to quantify, the total
uncertainty on $\log \epsilon(\mathrm{Li})$  is estimated to be up to
$\pm 0.6$\,dex, being larger for the cooler stars. The relative uncertainty
within the sample, however, is probably less than this.

\begin{figure*}
  \centering
  \includegraphics[width=\linewidth,bb=78 368 540 702]{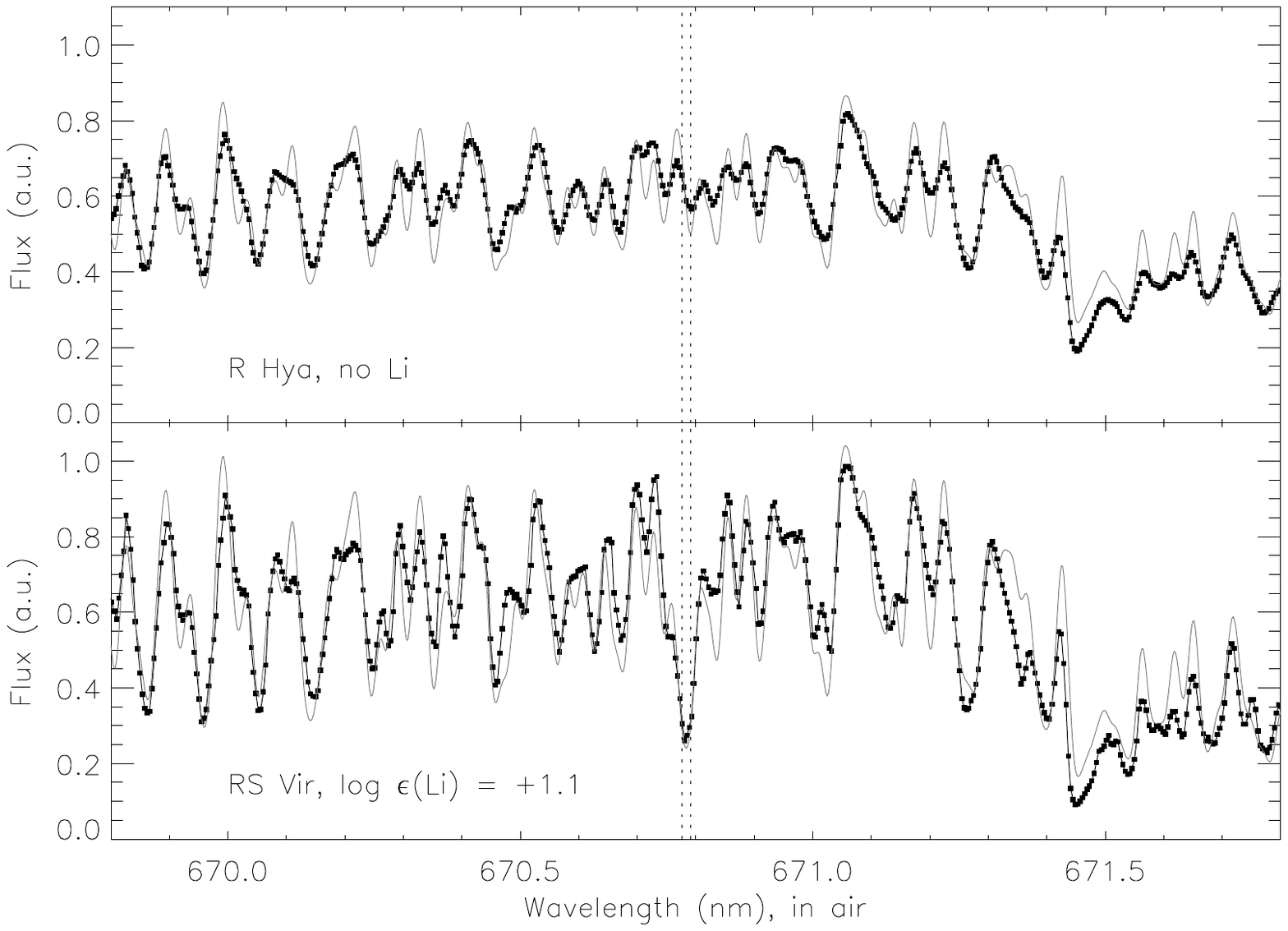}
  \caption{Examples of observed spectra (black graph with squares) of two stars
    with very different Li abundances, along with a best-fitting model spectrum
    (grey graph). The dashed lines indicate the laboratory wavelengths of the
    Li doublet.}
  \label{sampleLi}
\end{figure*}

A number of stars in the sample has no significant Li line absorption, their
spectra are satisfyingly fit with negligible Li abundance in the spectral
synthesis. It is difficult to state a precise upper limit to their atmospheric
Li abundance. In the vicinity of the Li line, positive and negative residuals
are present in an observed-minus-calculated flux plot, due to imperfect
spectral modelling. Besides errors on the adopted stellar parameters, neglecting
velocity fields in the atmosphere (due to stellar pulsations) by using
hydrostatic models leads to persistent observed-minus-calculated flux residuals.
Also, the line lists (mostly TiO) are certainly not without flaws at the
high resolution of our spectra. The stars with no detectable Li line absorption
are listed with $\log\epsilon(\rm{Li}) \lesssim 0.0$ in
Table~\ref{table_measure}, an abundance that would certainly not be detectable
with our methods anymore. The carbon-star \object{TW Hor} was analysed by
\citet{Kip04}, who determine a $\log\epsilon(\rm{Li}) = -1.5$ from another UVES
spectrum of this object. We adopt this value of the Li abundance for TW~Hor in
Table~\ref{table_measure}. We also compared our spectrum of TW~Hor with the one
used by \citet{Kip04} and found negligible variation between them. The low Li
abundance of the carbon-star TW~Hor would not be detectable in oxygen-rich
stars of the same effective temperature, because of the increased molecular
absorption in M-type stars in this spectral region. In the following we focus
on the oxygen-rich stars. Figure~\ref{sampleLi} shows sample spectra of two
stars (\object{R Hya} and \object{RS Vir}) with very different Li abundances,
along with a best-fitting model spectrum. The Li content of R~Hya is below the
detection threshold, whereas RS~Vir is found to have Li at the solar
photospheric level.

The very high Li abundance of \object{V441 Cyg} is remarkable. As noted above,
this star deviates from the trend towards higher (redder) $J - K$ colours at
lower temperatures: despite its rather high temperature estimate of 3400\,K, a
$J - K = 1.39$ is found. 
\citet{Van02} estimate
$T_{\rm{eff}} = 3100$\,K from its spectral type. This low temperature is hardly
compatible with our fit to the 700 -- 710\,nm spectral region. While there is
some disagreement on the near-IR photometry \citep{Wan02}, the combination of
red $J - K$ but weak TiO bands could be caused by a high C/O ratio. From our
models we found that a synthetic spectrum with C/O~=~0.9 and a temperature of
3200\,K also reasonably fits the 700 -- 710\,nm region. We thus repeated the Li
abundance measurement adopting C/O~=~0.9 and $T_{\rm{eff}} = 3200$\,K and
derived $\log\epsilon(\rm{Li}) = 4.1$, which is still a factor of 1000 higher
than the solar photospheric abundance! However, \citet{Ple03} demonstrate that
atomic and some molecular lines (such as from ZrO and LaO) strengthen even more
when the C/O ratio approaches 1. To find the Li abundance that may be regarded
as a lower limit, we constructed a MARCS model with a C/O ratio of 0.99 and
$T_{\rm{eff}} = 3400$\,K, but otherwise unaltered stellar parameters. Using
this model, we find $\log\epsilon(\rm{Li}) = 3.1$ for V441~Cyg. The TiO bands
are still well fit by the model at this high a C/O ratio.

The equivalent width of the Li\,I 670.8\,nm line in \object{V441 Cyg} is
approximately 730\,m\AA, surpassing that of many of the super-Li rich stars
identified in the Magellanic Clouds \citep{Smith95}, and the line core flux is
$\sim6\%$ of the local quasi-continuum level. Also, the Li line at 812.634\,nm
was identified in V441~Cyg. A spectrum synthesis of this line yielded a Li
abundance lower by 0.8\,dex (at solar C/O) than was found from the 670.8\,nm Li
line. However, the spectral fit in the vicinity of the 812.6\,nm line was not
as good as that around the 670.8\,nm line, which caused problems in defining
the quasi-continuum level.

We repeated the Li abundance measurement with the C/O ratio increased to 0.99
for \object{W Cet}, the other S star in our sample. This exercise yielded
$\log\epsilon(\rm{Li}) = -2.0$, which may again be regarded as a lower limit
for the Li abundance. The Li detection would still be significant, because the
detection threshold will also decrease markedly with C/O approaching 1. The
observed spectrum of W~Cet is still incompatible with negligible Li content, if
C/O~=~0.99 is adopted for the model atmosphere. For consistency, we keep the Li
abundances as determined with the C/O~=~0.48 models in
Table~\ref{table_measure} and in the analysis.

We would also like to note that we did not identify significant LaO bands in
any of our oxygen-rich stars, thus we may assume that indeed none of them has a
C/O ratio above 0.99.

Besides V441~Cyg, only one of our sample stars has previously been investigated
for its Li content, namely \object{RS Vir} \citep{GH07}. In contrast to our
measurement (Table~\ref{table_measure}), \citet{GH07} find no significant Li
line absorption in that star, and state an upper limit of
$\log\epsilon(\rm{Li}) < 0.5$ instead. Also, the temperature determinations of
this star differ somewhat, 2900\,K in the present work compared to 2700\,K in
\citet{GH07}. We identify two possible explanations for this discrepancy.

First, two different TiO line lists were applied to derive stellar
temperatures. One of them might yield systematically lower temperatures, which
would decrease the measured Li abundance considerably. To investigate this
possibility, we calculated model spectra using the TiO line list of
\citet{Ple98}, as used by \citet{GH07}, and repeated the $\chi^2$ minimisation
procedure (Sect.~\ref{sect_param}). Only the sequence of models with solar
metallicities was used for this. From this we find that the temperatures found
using the two TiO line lists hardly differ from each other, with the Plez list
delivering temperatures at most 50\,K {\em higher} than the Schwenke list. We
conclude that the different line lists used cannot explain the discrepancy
found in the temperature and Li content of \object{RS Vir}. From this
experiment with the two TiO line lists, we also find that the list of
\citet{Sch98} yields slightly lower residuals than that of \citet{Ple98},
particularly in the vicinity of the 670.8\,nm Li line. This might also explain
why our detection limit is somewhat lower than that of \citet{GH07}.

Second, the two spectra have been taken at different epochs, with different
instruments. The temperature variation over the cycle of stellar pulsation
could be responsible for the difference of 200\,K found by the two studies.
According to the AAVSO visual light curve, our spectrum was taken just 30 days
before maximum, when the light curve was still sharply rising. \citet{GH07}
observed \object{RS Vir} during their fourth run between 21 -- 24 February
1997. This date is just a few tens of days before a visual light maximum of the
AAVSO light curve, too, but three pulsation cycles before our UVES
observations. With the pulsational phase at the two observations being so
similar, it is hard to imagine that the photospheric temperature could be
different by 200\,K, even for a Mira variable. Assuming a Li abundance constant
on short time-scales, it is very unlikely that the line would go undetected in
the observations of \citet{GH07}. The atmospheric structure of a Mira star
certainly cannot be described perfectly with a hydrostatic model, and line
strengths might vary considerably with the pulsation phase, even from cycle to
cycle. It would be interesting to obtain another spectrum of RS~Vir to exclude
a variable Li line strength, as has been reported for a number of carbon-stars
by \citet{Bof93}.

A portion of the spectrum of \object{RS Vir} around the 670.8\,nm Li line is
displayed in Fig.~\ref{sampleLi}, with the Li line clearly present.

\section{Discussion}\label{discussion}

\begin{figure}
  \centering
  \includegraphics[width=\linewidth,bb=6 1 537 388]{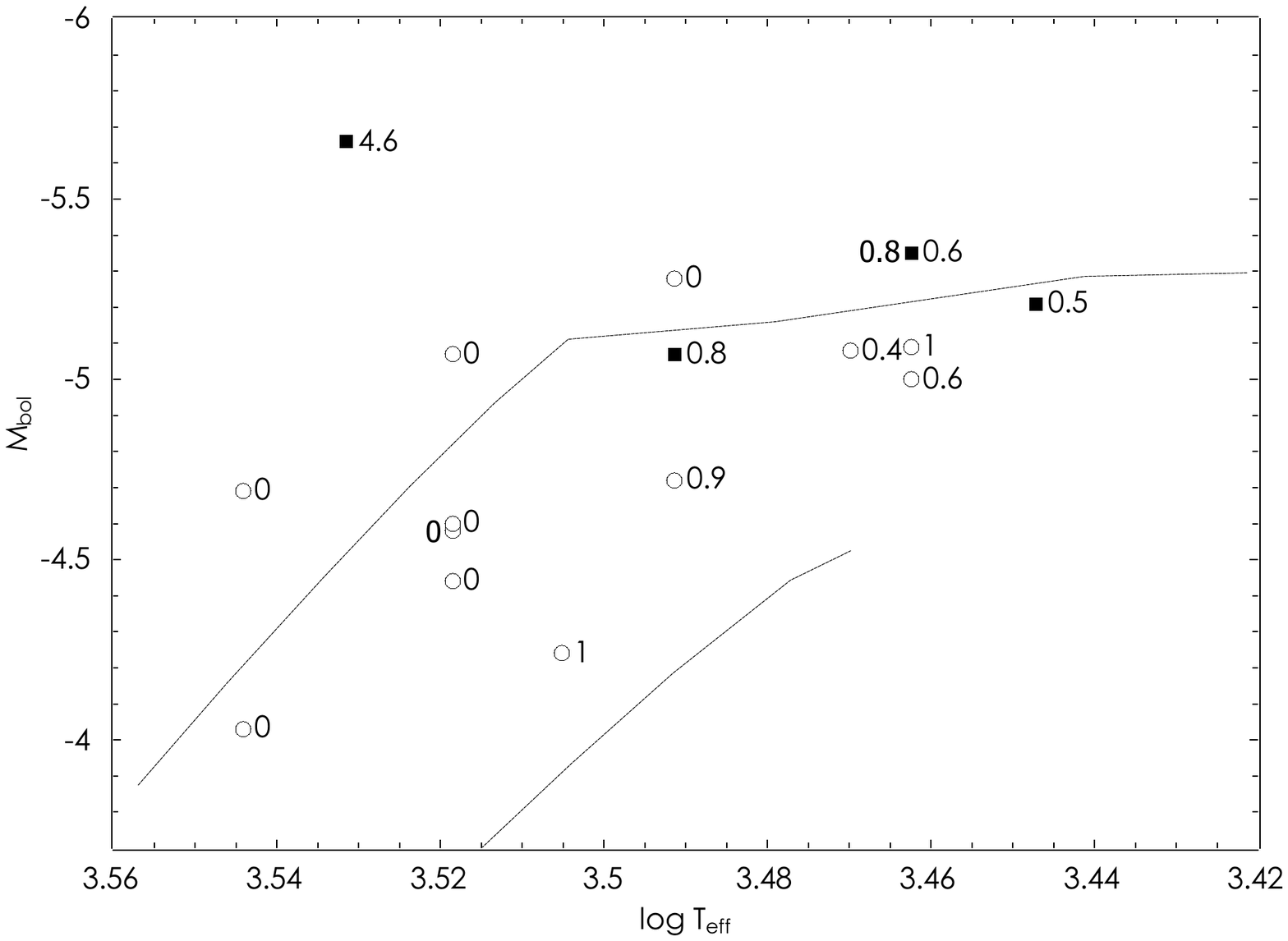}
  \caption{$T_{\rm{eff}}$ vs.\ $M_{\rm{bol}}$ diagram for our sample stars.
    Filled and open symbols denote objects with and without Tc, respectively.
    The stars S~Gru and R~Hor overlap at
    $(M_{\rm{bol}}, \log T_{\rm{eff}}) = (-5\fm35, 3.46)$. The C-star TW~Hor
    has been excuded from this plot. The labels give the Li abundance derived
    (see text). The solid lines are parts of isochrones of 400\,Myrs age
    (corresponding to an AGB mass of $\sim3\,\rm{M}_{\sun}$) and 6.3\,Gyrs (AGB
    mass $\sim1.05\,\rm{M}_{\sun}$), respectively, with solar metallicity and
    no dust formation from \citet{Marigo08}. The isochrones have been cut off
    once the model stars become C-rich.}
  \label{HRD}
\end{figure}

In Fig.~\ref{HRD} we show the location of our stars in a $T_{\rm{eff}}$
vs.\ $M_{\rm{bol}}$ diagram using the data given in Table~\ref{table_charac},
along with two isochrones from \citet{Marigo08}. Most stars nicely follow a
trend of increasing luminosity with cooler temperature (higher $J-K$). Stars
with and without Tc are marked by different symbols. In agreement with earlier
studies \citep[e.g.][]{LH03,Utt07a}, we find no Tc-rich stars below
$M_{\rm{bol}}=-4.8$. This is another confirmation of the lower luminosity limit
for third dredge up predicted by various stellar evolution models. As in
earlier studies, we also find here a few stars without Tc above the luminosity
limit. We refer to the discussion in \citet{LH03} that suggests that these
stars might have too low an envelope mass for third dredge up to occur.

\begin{figure}
  \centering
  \includegraphics[width=\linewidth,bb=6 4 539 382]{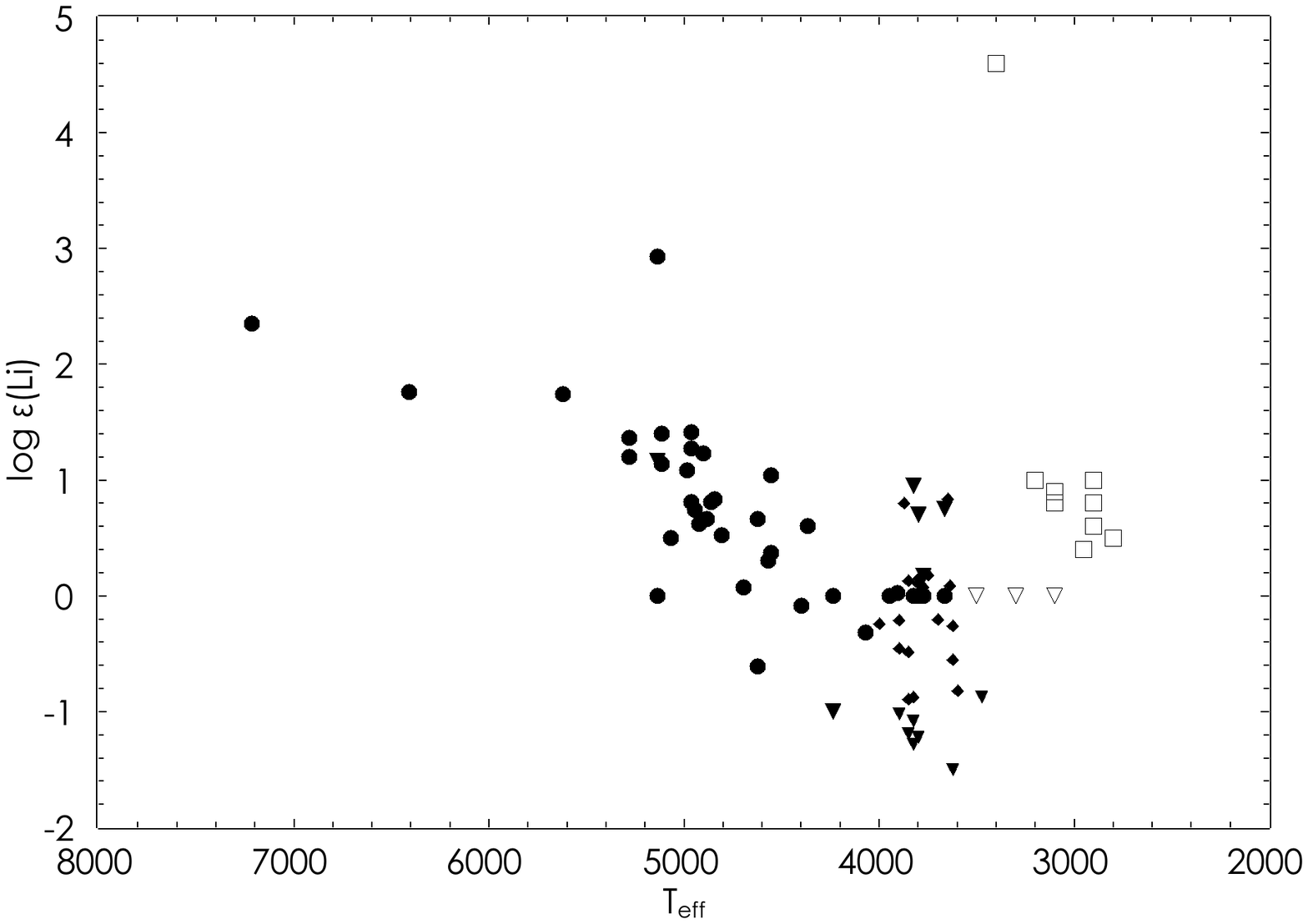}
  \caption{Li abundance against effective temperature for the stars in our
    sample (open symbols) and the giants from the list of
    \citet[][filled diamonds]{Luc82} and \citet[][filled circles]{Mallik99}.
    Large triangles represent upper limits from the sample of Mallik, small
    triangles those from Luck \& Lambert. The average Galactic value for
    the Li abundance is $\log \epsilon(\rm{Li}) = 3.1$ \citep{Michaud91}.}
  \label{LivsT}
\end{figure}

The data points in Fig.~\ref{HRD} are labelled with the derived Li-abundances
from Table~\ref{table_measure}. Standard theories predict a Li abundance on the
AGB below $\log \epsilon(\rm{Li}) = 1.5$. Only one star in our sample exceeds
this value, namely \object{V441 Cyg}. We discuss this star in more detail in
Sect.~\ref{sec_V441Cyg}. For the other stars, there seems to be a tendency to
higher Li abundance with higher luminosity or with lower temperature. This
contradicts the observations of giants summarised in \citet{Michaud91} and
\citet{Mallik99}, as they find a decrease in the average Li abundance towards
lower temperature. In Fig.~\ref{LivsT} we show our sample, together with the
results for the giants presented by \citet{Luc82} and \citet{Mallik99} in a
$T_{\rm{eff}}$ vs.\ $\log \epsilon(\rm{Li})$ plot. Our sample extends the
previous dataset towards cooler temperatures. This extension reveals that the
trend to decreasing Li abundance with lower temperature stops and may even be
reversed at the cool end.

The decline in the Li abundance as a function of temperature is typically
attributed to two effects \citep{Mallik99}: first, the surface abundance will
be reduced by a deepening of the convective envelope as the stars evolve
towards and up the giant branch. Second, mass loss should erode outer Li-rich
layers during the advanced stages of stellar evolution.

At the lowest luminosities we find in our sample only stars with a strong
depletion of their Li abundance (see Fig.~\ref{HRD}). These stars also have the
highest temperature, thus link our sample to the one of \citet{Mallik99}. A
deep extension of the convective envelope results in the destruction of most of
the Li. Lines of this element again appear in the spectrum at about
$M_{\rm{bol}}=-4.7$, i.e.\ at a slightly lower luminosity than the threshold
level for Tc occurrence. This suggests that production of Li sets in at some
point. This is supported by Li preferentially being detected in sample
stars with long pulsation periods ($P \gtrsim 280$\,d) and redder IRAS
[12] -- [25] colours. Based on the derived luminosities, HBB may be safely
excluded as the production mechanism for all but one star.

\citet{Utt07b} found four AGB stars in the Galactic bulge with a detectable Li
abundance of $\log \epsilon(\rm{Li})$ between 0.8 and 2.0 (at an error of 0.4
dex) and an effective temperature around 3000\,K. These results agree nicely
with our findings. Uttenthaler et al.\ identify extra-mixing
\citep[e.g.][]{bs99} as the most likely explanation for the re-appearance of Li
on the stellar surface. We speculate that the same mechanism is at work in the
Li-enhanced\footnote{We use here the term `enhanced` to express a Li abundance
above our detection threshold corresponding to $\log \epsilon(\rm{Li})=0$. This
has to be distinguished from {\it Li rich} objects with an abundance exceeding
the average Galactic value.} stars of our sample. The lowest luminosity of a
Li-enhanced star in the sample of \citet{Utt07b} is in good agreement with the
lower luminosity limit found among the targets studied here.

In this sample we cannot distinguish whether this change in Li abundance is
coming from a difference in mass or as a result of stellar evolution or both.
However, when combining our findings with the ones from \citet{Utt07b}, there
seems to be a lower luminosity limit for any Li production mechanism to set in.
The existence of such a limit would favour an internal mechanism over an
external one
\citep[e.g.\ surface contamination by a companion -- see][]{Utt07b}. We note
further that mass loss increases substantially during the AGB phase. As
mass loss should reduce the surface Li abundance, the detection of an increase
in Li abundance on the upper AGB strongly suggests a production of this element
inside the star.

Concerning a possible relation between the Li abundance and the occurrence of
Tc, we find that all Tc-rich stars show also at least a mild enrichment of Li
relative to the values found for the lowest luminosity stars of our sample. We
also find Li-enhanced stars without Tc and luminous stars that show no
detectable abundance of neither of the two elements. Furthermore, \citet{Utt07b}
also found Tc rich stars with no Li-enhancement. This mixture of possibilities
indicates the involvement of a further parameter besides the stellar
luminosity. As we have no indications of a strong scatter in metallicity in our
sample, we may assume that mass is the discriminating factor. There are some
indications that extra-mixing occurs more effectively among stars of lower
mass, which may lead to an enhancement of surface Li in these objects
\citep[e.g.][]{Can08}.

If we want to compare our results with the work of \citet{Van07}, who find
objects with all four combinations of Tc-rich, Tc-poor, Li-rich, and Li-poor,
we face the problem that Vanture et al.\ define stars as Li-rich if
$\log \epsilon(\rm{Li}) \ge 1.0$. All other stars are defined as Li-poor. With
this approach, only two stars, \object{RS Vir} and \object{V441 Cyg} would be
classified as Li rich, with the first one at the lower limit. In that sense we
would find one Tc-rich/Li-rich star (see below), one Tc-poor/Li-rich star, four
(five if we include TW Hor) Tc-rich/Li-poor stars and a couple of
Tc-poor/Li-poor stars. This means that all four classes from \citet{Van07}
would be present in our sample. However, taking into account the uncertainties
in the Li abundance and the obvious scatter of Li abundances between
$\log \epsilon(\rm{Li})=0$ and 1, we think that the Vanture et al.\ classes
oversimplify the picture. Our understanding would certainly profit from finding
an abundance limit between mildly Li-enhanced low-mass AGB stars and luminous
Li-rich AGB stars producing this element by HBB in order to disentangle two
different production mechanisms.

That all four combinations of Tc-rich, Tc-poor, Li-rich, and Li-poor
are found can be understood at least qualitatively from the theoretical point
of view. Assuming that $^3$He is left from the RGB phase, thermohaline mixing
\citep{Can08} can connect the H-burning shell of a TP-AGB star with the
convective envelope and thus transport $^7$Be to cool layers where it may decay
to Li that can be observed on the surface. This mechanism is found to be
more efficient in low-mass stars \citep[$\simeq1.5$\,M$_{\sun}$,][]{Can08}.
This scenario is not constrained to stars undergoing 3DUP events, although the
faster mixing during 3DUP would ensure that even more $^7$Be survives. In
higher mass stars ($\gtrsim3$\,M$_{\sun}$), thermohaline mixing cannot connect
the H-burning shell with the convective envelope. Nevertheless, a $^7$Be rich
pocket can build up above the H-burning shell that can be dredged to the
surface after a TP, even before formal 3DUP occurs. The $^7$Be-rich pocket
would be dredged to the surface in any 3DUP event, which may account for our
finding of Li in every Tc-rich star in the current sample. The requirement that
some $^3$He is preserved from the RGB phase also introduces a mass threshold
above which extra mixing can work on the AGB, because for stars of too low an
initial mass, all $^3$He would be consumed by efficient extra-mixing already on
the RGB. Thus, thermohaline mixing critically depends on the mixing history of
the star.

In the magnetic buoyancy scenario \citep{Gua09}, the mixing is thought to arise
from magnetic bubbles with locally increased magnetic pressure, thus a decreased
density. The magnetic field might be sustained by an $\alpha - \Omega$ dynamo
produced by a region of rotational shear below the convective envelope. Fields
of different strengths descending from different initial rotational velocities
may explain the observed spread in Li abundances. Independent from 3DUP, Li
may be produced on the AGB by fast mixing rates (provided a $^3$He reservoir is
still present from the RGB), or destroyed by slow mixing rates.

Quantitative predictions of models of extra mixing are still rather
uncertain and require parameter fine-tuning by observations. Depending on the
adopted efficiency parameter of thermohaline mixing, \citet{Can08} predict a
surface Li enrichment of $\sim0.05$\,dex during one inter-pulse period of a
3\,M$_{\sun}$ TP-AGB model star. After a number of 3DUP events, the surface Li
abundance may thus increase above the detection threshold in a cool atmosphere.
Thermohaline mixing is expected to be more efficient in lower mass stars, so
the surface Li abundance could be increased by larger amounts than for the
described case of a 3\,M$_{\sun}$ star. It should be noted that the interaction
of thermohaline mixing with magneto-rotational instabilities is important, but
only full magneto-hydrodynamical simulations can lead to a better understanding
of these interactions.

The magnetic buoyancy picture, on the other hand, predicts an upper limit
on Li production at the level of $\log \epsilon(\rm{Li}) = 2 -2.5$. This upper
limit is set by assuming a short phase of very fast mixing (a few km\,s$^{-1}$)
of magnetised bubbles, in which no burning of the material occurs, and the Li
enrichment of the envelope is set by the amount of ``normal'' $^7$Be that is
produced near the H-burning shell. The Li abundances below this upper limit can
be easily reproduced by assuming some Li destruction by a slower transport of
the material, such that some nuclear burning occurs during the transfer to the
envelope. If magnetic fields are the origin of the extra mixing, both fast and
slow regimes of circulation can explain very different outcomes with only a
single physical mechanism.

To conclude, super-Li rich stars such as \object{V441 Cyg} cannot be
easily explained by the magnetic buoyancy scenario, but the rest of our sample
would be described well. To decide whether thermohaline mixing could be the
dominant mechanism, more information on the masses and mixing histories of the
stars would be required.

\smallskip
We end this discussion with a closer look onto two of our sample stars.
As far as we know, the reclassification of \object{W Eri} as Tc-poor has no
great consequences on understanding the nature of this star. However, the
re-classification of \object{V441 Cyg} and \object{R Hya} as Tc-rich and
Tc-poor, respectively, has important consequences, and we thus devote the
following two sections to these two objects.

\subsection{V441 Cygni}\label{sec_V441Cyg}

\begin{figure}
  \centering
  \includegraphics[width=8.5cm,bb=73 387 541 704]{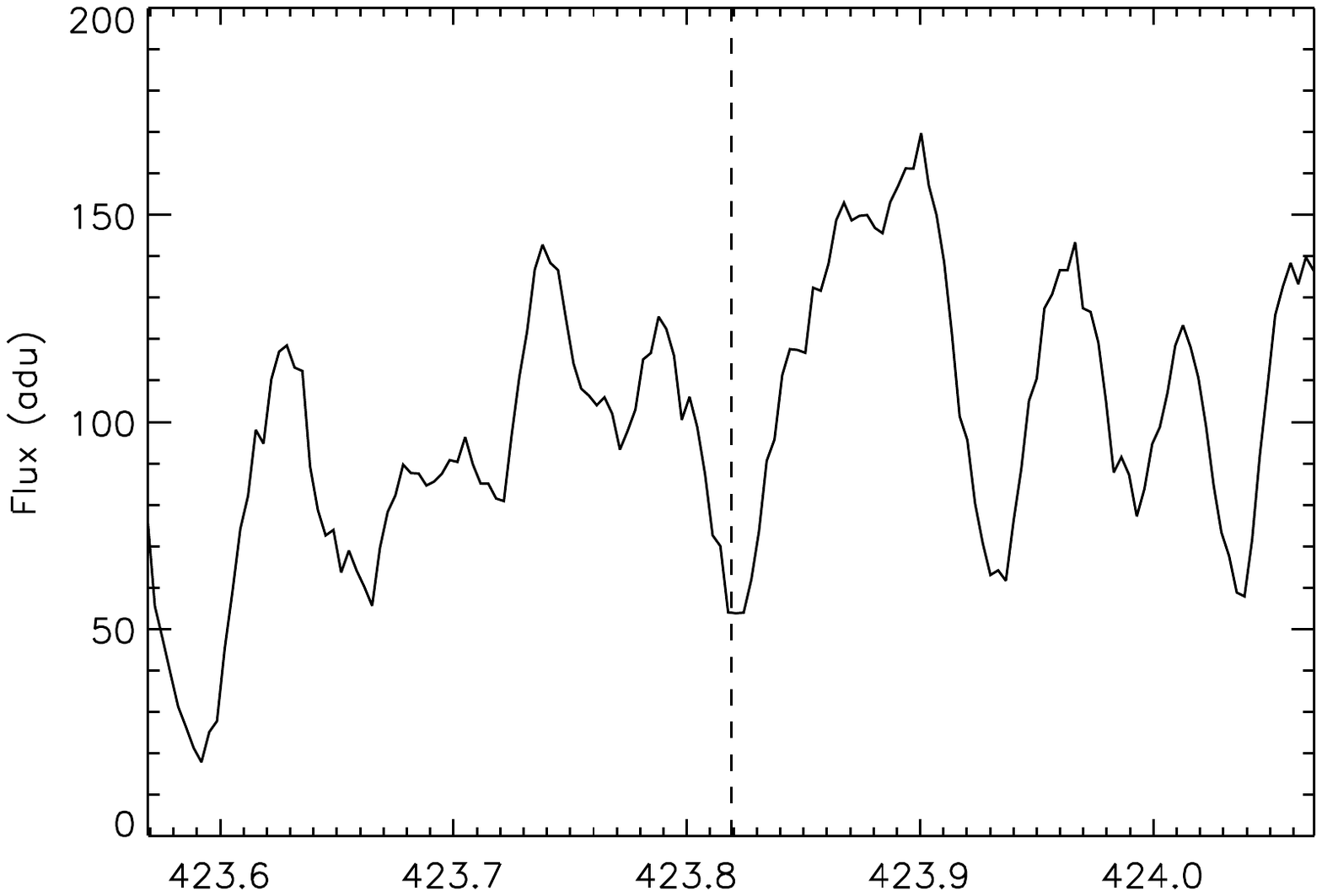}
  \includegraphics[width=8.5cm,bb=73 387 541 704]{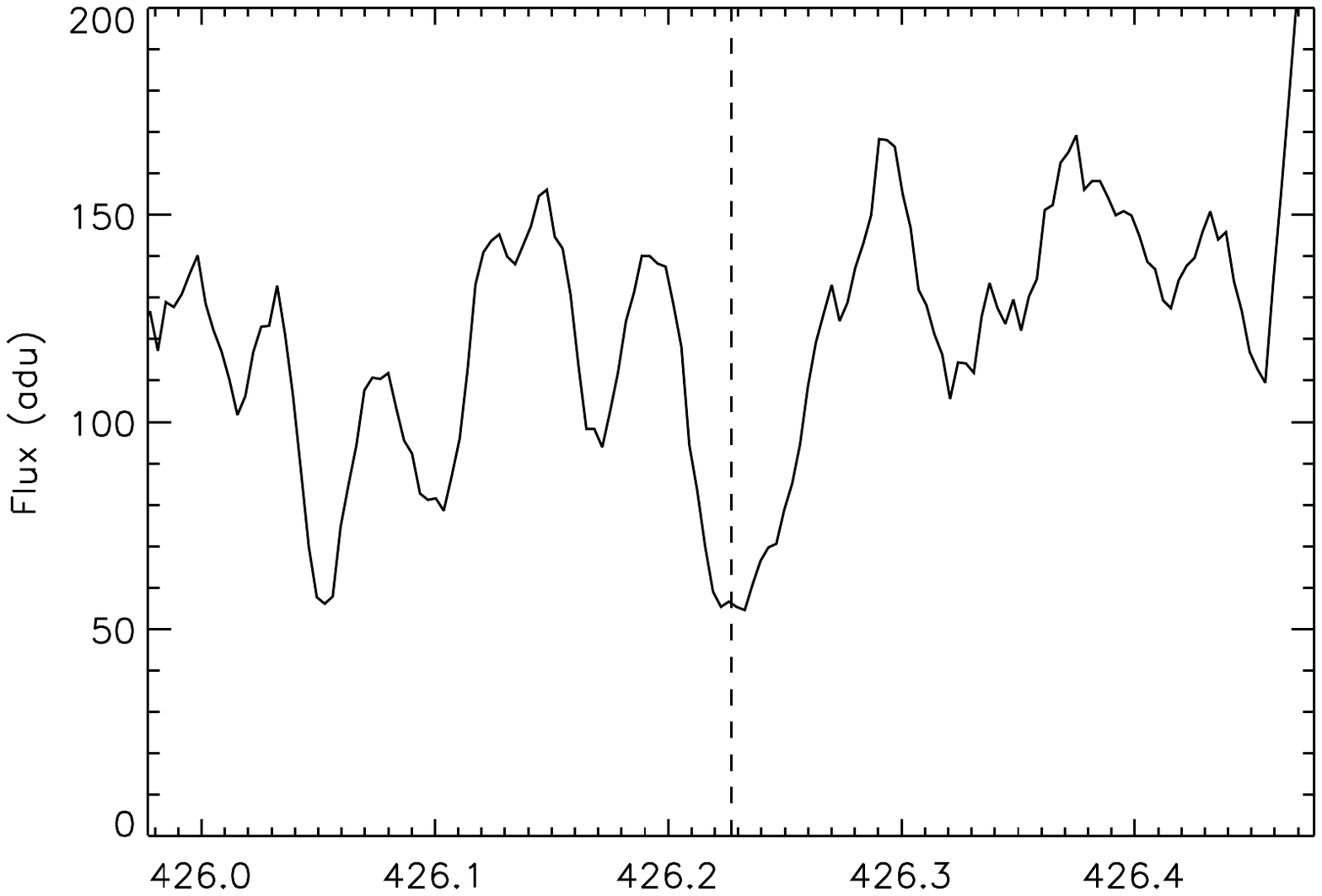}
  \includegraphics[width=8.5cm,bb=73 368 541 704]{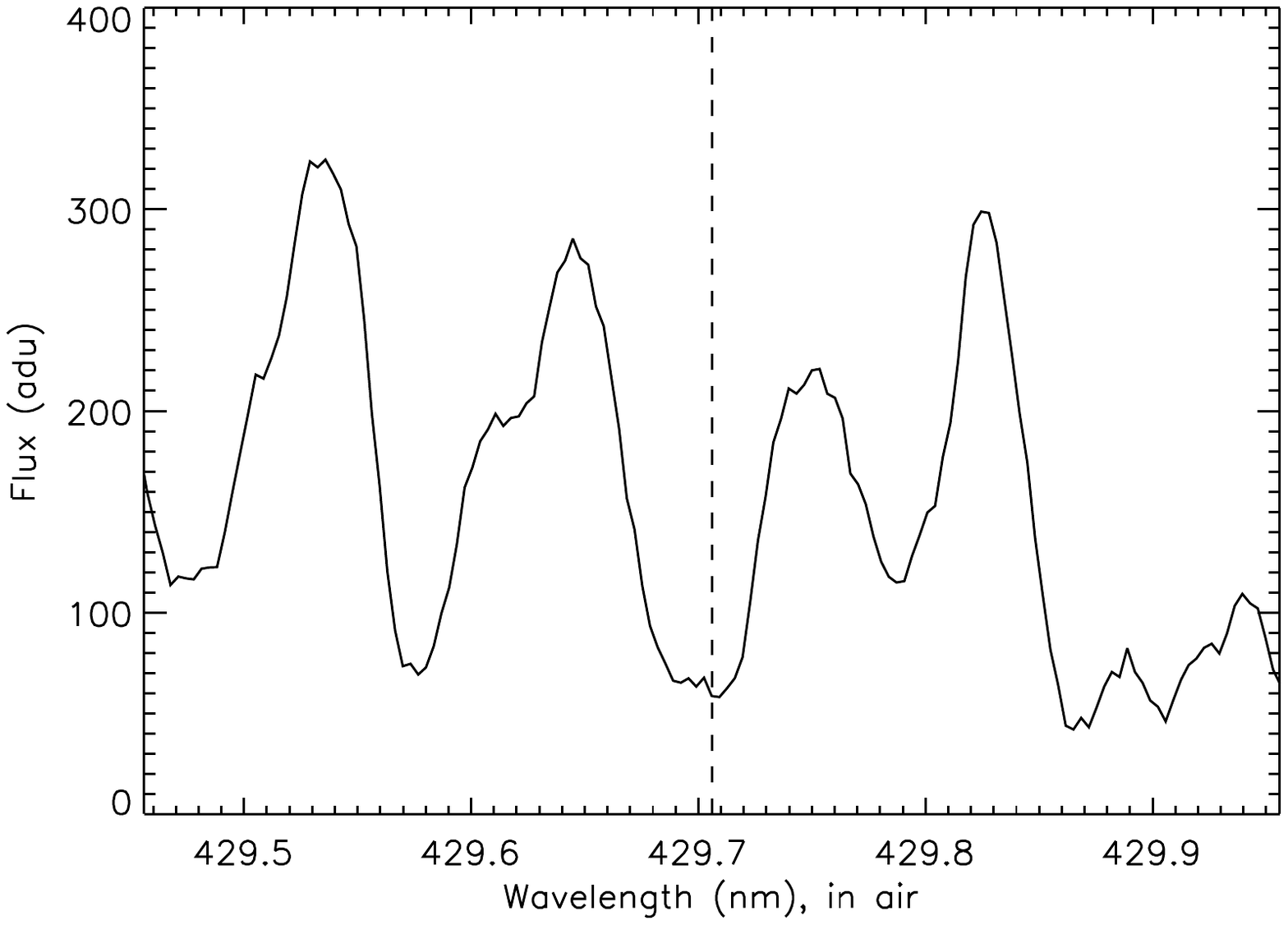}
  \caption{FOCES spectrum of \object{V441 Cyg} around the three strongest TcI
    lines. The spectrum has been smoothed with a 5 pixel boxcar. The dashed
    line marks the laboratory wavelength of the respective line. Clearly, all
    three TcI lines are present.}
  \label{lines}
\end{figure}

Even though the SNR of the spectrum of \object{V441 Cyg} is low, it is clear
from Fig.~\ref{line_ratio} that this star indeed has Tc in its atmosphere. It
is thus an intrinsic S-star. As further evidence, its spectrum around the three
strongest Tc lines is shown in Fig.~\ref{lines}. Also \citet{Wan02} list
V441~Cyg as ``{\em not} a good candidate Tc-deficient S star'' based on its
near- and mid-IR photometry.

\citet{Van07} investigate a sample of 59 Galactic S-stars for the correlation
of Tc and Li and find all four possible combinations of Tc-rich and -poor, and
Li-rich and -poor in their sample. However, the group of Tc-poor, Li-rich
S-stars is formed by only one object, namely \object{V441 Cyg}. \citet{Van07}
thus call it a ``unique'' object. There was good reason for doubt because
V441~Cyg actually has never been checked for its Tc content in the literature,
hence our FOCES observations. As shown above, there is no Li-rich (not
{\em Li-enhanced}!) and Tc-poor S-type star known at the moment.

We thus show that \object{V441 Cyg} is Tc- and Li-rich and that it needs to be
moved to ``group 2'' as defined by \citet{Van07}. They offer two possibilities
for interpreting this group of stars. The first interpretation is that
intrinsic (Tc-rich) S-stars with high Li-abundance are intermediate-mass
($M \gtrsim 4\,\rm{M}_{\sun}$) TP-AGB stars experiencing HBB, thus producing Li
in their atmospheres and destroying C via CN cycling. This picture is supported
by the low average Galactic latitude of the stars in group 2 (14$\degr$,
excluding two relatively nearby stars). However, the bolometric magnitude,
which is available for only four stars of the \citet{Van07} sample, instead
argues against this interpretation, since none of them has
$M_{\rm{bol}} \le -6\fm0$. \citet{Van07} thus suggest that their sample
consists of a mix of low- and intermediate-mass stars. The second
interpretation thus concerns low-mass S-stars, which might be the result of
extra-mixing processes on the TP-AGB (see discussion above). The existence of
Li-rich C-stars \citep{Bof93} suggests that low-mass Li-rich S-stars could also
exist, and have not yet dredged-up enough carbon to reach C/O~$>$~1.

Which of the two offered interpretations holds for \object{V441 Cyg}? It is
located at a very low Galactic latitude of $b\sim-1\degr$, and our estimate of
its bolometric magnitude ($M_{\rm{bol}} = -5\fm66$) is not far below the
approximate limit for HBB. Concerning the uncertainties in determining the
luminosity we note that only if the star were be a fundamental mode pulsator,
we would expect a lower luminosity. This definitely argues in favour of the
intermediate-mass HBB interpretation. However, \citet{GH07} find that Li-rich
Galactic OH/IR stars, thought to be intermediate-mass AGB stars with a very
high mass-loss rate, do not exhibit significant Zr enhancements. On the other
hand, the s-element Rb is observed to be enriched in these stars \citep{GH06}.
This might stem from the activation of the $^{22}$Ne($\alpha$,n)$^{25}$Mg
neutron source rather than the $^{13}$C($\alpha$,n)$^{16}$O neutron source in
these more massive stars. Thus, enrichment in Rb is another indicator of high
mass. Unfortunately, we did not succeed to derive a reliable Rb abundance from
our spectrum of V441~Cyg, because the Rb resonance line at 780.0\,nm is
completely dominated by a blue-shifted circumstellar component that makes it
impossible to determine the photospheric Rb abundance. Nevertheless, we can
interpret this as no enhancement of Rb in the photosphere of V441~Cyg,
suggesting that it is too low in mass to activate the $^{22}$Ne neutron source.
The light curve of V441~Cyg changes with a period of 288\,d in a semi-regular
way, and the IRAS [12] -- [25] colour, defined as $-2.5 \log(F_{12}/F_{25})$,
has a value of $-1.28$. Only three of our sample stars are bluer than that,
which suggests only low obscuration by the circumstellar envelope. The
variability characteristics, in particular the low amplitude of pulsation at
this high luminosity, could be interpreted as a consequence of a high stellar
mass, but is difficult to quantify.

To summarise, the evidence on the mass of \object{V441 Cyg} is not entirely
conclusive, and we cannot clearly decide on which of the interpretations
offered by \citet{Van07} holds for this star. There are a number of other
peculiar intrinsic S-stars with high Li abundance, e.g.\ \object{R And} and
\object{T Sgr}. The Rb abundance has been measured in the former by
\citet{Wal92}, and a neutron density at the s-process site of 10$^8$\,cm$^{-1}$
was derived, indicative for the $^{13}$C neutron source operating in low-mass
stars. An analysis and comparison of these stars would be be very rewarding,
which we plan to present in a forthcoming paper.

\begin{figure}
  \centering
  \includegraphics[width=\linewidth,bb=82 370 553 700]{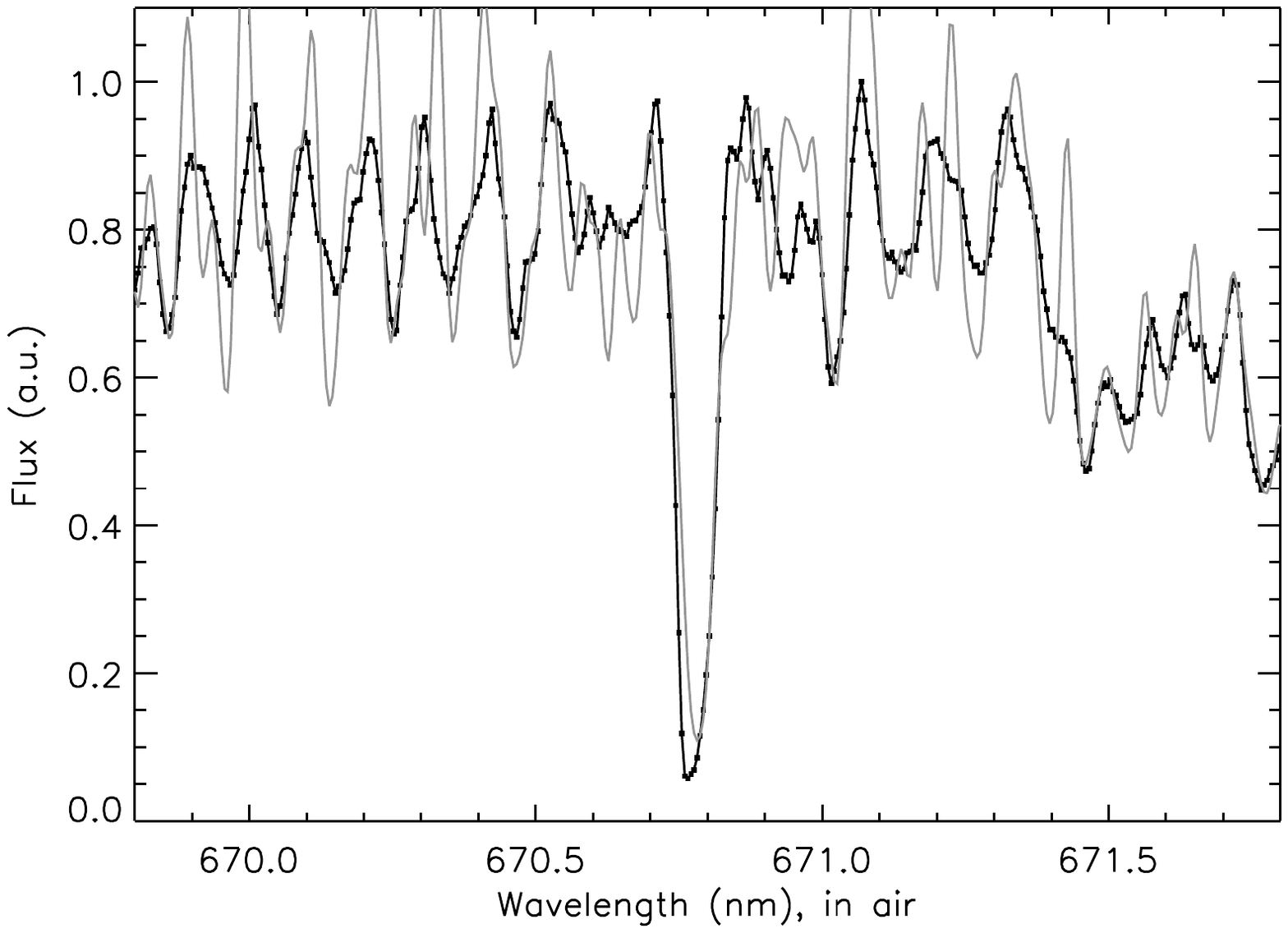}
  \caption{FOCES spectrum of \object{V441 Cyg} around the 670.8\,nm Li
    resonance line (black graph with squares), and a best fitting synthetic
    spectrum (grey graph) based on a MARCS model atmosphere with
    $T_{\rm{eff}} = 3400$\,K and solar metallicity. A Li abundance of
    $\log \epsilon(\rm{Li}) \simeq +4.6$ has been assumed in the spectral
    synthesis. We did not try to fit the blue wing of the Li line, because we
    interpret it to be caused by a Li-rich stellar wind from V441~Cyg.}
  \label{V441CygLiline}
\end{figure}

Finally, Fig.~\ref{V441CygLiline} reproduces a small piece of spectrum of
\object{V441 Cyg} around the 670.8\,nm Li resonance line, along with the
best-fitting synthetic spectrum based on a MARCS model atmosphere with
$T_{\rm{eff}} = 3400$\,K and solar metallicity, assuming
$\log \epsilon(\rm{Li}) \simeq +4.6$. There is a prominent blue wing in the Li
line, which we did not try to fit. This feature may be interpreted as indication
of a Li-rich stellar wind that cannot be reproduced with our hydrostatic models.
Such a wind feature is somewhat surprising, given the rather blue IRAS colours
of this star. Because the temperature is probably lower in the wind than in the
photosphere, more Li atoms are present in the ground state, and the wing is
deeper than the photospheric line component. Thus, we possibly see V441~Cyg in
a stage where it enriches the ISM with freshly produced Li.

\subsection{R Hya}

\object{R Hya} has already been classified as Tc-rich a long time ago, albeit
with only a low Tc line intensity \citep{Mer52b}. Spectrogram tracings of
R~Hya taken at different phases of its pulsation cycle can be found in
\citet{Mer52a}. These spectrograms cover the Tc\,I lines at 398.497 and
404.911\,nm, but both are absent at all observed pulsational phases. Also in
our high-quality, high-resolution UVES spectra of R~Hya, we are unable to
detect any of the Tc lines. We find no increase in ZrO band strength, wether in
R~Hya or in any other Tc-poor star in our sample. We thus conclude that, with
the spectrograms available at the times of Merrill, a precise decision about
the presence of weak absorption lines was not possible, and a misidentification
occurred in this particular case that has continued in the literature ever
since. We find R~Hya to be the most luminous Tc-poor star in our sample, with
approximately 10\,000 solar luminosities. Because of its Mira type variability
with a rather long pulsation period, as well as its slightly redder IRAS
colours, it is probably in a more advanced AGB stage than other suspected truly
low-mass AGB stars, such as \object{ER Vir} and \object{EV Vir}.

\object{R Hya} is of particular interest because its pulsation period has
strongly decreased in time \citep{Tem05}. It is thus speculated that it has
recently undergone a TP \citep{WZ81}. If this interpretation of the period
change is correct, then either the most recent thermal pulse was not followed
by a third dredge-up event, or the dredged-up inter-shell material has not yet
reached the surface of the star. The absence of Tc in the atmosphere of R~Hya,
however, does not exclude the possibility that a recent TP is indeed the cause
of the observed period decrease.

From detailed AGB evolution models \citep[][]{Cri09}, it is known that a 3DUP
episode, if it takes place, follows a TP with some time delay. For instance, for
a 2\,M$_{\sun}$ model of solar metallicity, the first 3DUP episode takes place
$\sim670$ years after the preceeding TP (S.\ Cristallo, private communication).
Thus, abundance changes of s-process elements and $^{12}$C on the stellar
surface also become visible only with a time delay. \citet{WZ81} suggest that a
TP could have occurred in \object{R Hya} about 550 years ago. If the decrease
in pulsation period observed in R~Hya was indeed caused by a recent TP, we may
be able to detect surface abundance changes (e.g.\ Tc) in the future, say in a
few 100 years, if this past TP is followed by a 3DUP event.\footnote{For the
s-process elements to be enhanced, it takes at least two 3DUP episodes, one to
build up a $^{13}$C pocket in the inter-shell region as neutron source, and
another one to actually mix the s-process products to the surface. C
enhancement should be present after only one 3DUP episode.}

Alternatively, \citet{Zij02} suggest that the period decrease observed in
\object{R Hya} could just as well be caused by a nonlinear instability leading
to an internal relaxation of the stellar structure. While not detecting Tc does
not strongly favour one or the other scenario, it indicates that possible TPs
are not strong enough to drive 3DUP in R~Hya.

\section{Summary and conclusions}\label{summary}

We analysed high-resolution optical spectra of a sample of 17 M-type
(oxygen-rich) and one C-type (carbon-rich) AGB variables for the presence of Tc
and the abundance of Li. We employed a flux-ratio method to decide on the
presence of the 3$^{\rm{rd}}$ dredge-up indicator Tc in the stellar atmosphere,
and measured the abundance of the temperature-sensitive element Li with
spectral synthesis techniques of the 670.8\,nm Li resonance line using MARCS
model atmospheres. Bolometric magnitudes were established from near-IR
photometry and pulsation periods. The lower luminosity limit for 3DUP to occur,
established in previous works, has been confirmed. We re-classified three
sample stars with respect to their Tc content, namely that \object{R Hya} and
\object{W Eri} are Tc-poor, and \object{V441 Cyg} is Tc-rich. We thus suggest
that R~Hya has not yet experienced a dredge-up of freshly produced s-elements,
but, regarding the possibility that it recently underwent a TP, might have a
3DUP episode in the future. V441~Cyg, on the other hand, is an intrinsic
S-star.

Lithium is detected in ten sample stars (with a $\log \epsilon(\rm{Li}) > 0$).
After combining these abundances with bolometric magnitude estimates, we find
that Li is more abundant in more luminous and cool stars. All Tc-rich sample
stars are found to contain also some Li in their atmosphere. We thus suggest
that this fragile element could be replenished in the stellar atmosphere by
some (extra-mixing) process that is only active in the more luminous objects.
This could also be connected to a mass effect within the sample.
The S-star \object{V441 Cyg} is found to be super-Li rich, with a higher Li
abundance than the solar photospheric value probably by a factor 1000. The
gathered evidence does not allow us to clearly conclude whether this object is
an intermediate-mass AGB star ($M \gtrsim 4\,\rm{M}_{\sun}$) producing Li by
the HBB mechanism, or a low-mass star with extremely efficient extra-mixing
processes. The data obtained in this study may be useful for constraining
theoretical models of such extra-mixing processes.

\begin{acknowledgements}
  We thank M.\ Busso and M.\ Cantiello for their fruitful comments, M.\ T.\
  Lederer for constructing the model atmospheres used in this study, J.\ Hron
  for carrying out the visitor-mode observations with UVES/VLT, the Calar
  Alto Observatory staff for carrying out the service mode observations of
  V441~Cyg, and the anonymous referee for useful comments. SU acknowledges
  support from the Fund for Scientific Research of Flanders (FWO) under grant
  number G.0470.07. TL acknowledges support by the Austrian Science Fund FWF
  under project number P20046-N16. This publication makes use of data products
  from the Two Micron All Sky Survey, which is a joint project of the
  University of Massachusetts and the Infrared Processing and Analysis
  Center/California Institute of Technology, funded by the National Aeronautics
  and Space Administration and the National Science Foundation. We acknowledge
  with thanks the variable star observations from the AAVSO International
  Database contributed by observers worldwide and used in this research.
\end{acknowledgements}

\end{document}